\documentclass[final,5p,times,twocolumn]{elsarticle}	
 \usepackage{lineno}
 \usepackage{color}

\usepackage{amssymb}
\usepackage{subfigure}
\begin{document}
\begin{frontmatter}
\title{Progress in coupling MPGD-based Photon Detectors with Nanodiamond Photocathodes}

\author[f]{F.~M.~Brunbauer}
\author[a]{C.~Chatterjee}
\author[d]{G.~Cicala}
\author[e]{A.~Cicuttin}
\author[e]{M.~L.~Crespo}
\author[b]{D.~D`Ago}
\author[a]{S.~Dalla~Torre}
\author[a]{S.~Dasgupta}
\author[a]{M.~Gregori}
\author[a]{S.~Levorato}
\author[c]{T.~Ligonzo}
\author[f,g]{M.~Lisowska}
\author[d]{M.~S.~Leone}
\author[e]{R.~Rai}
\author[f]{L.~Ropelewski}
\author[a]{F.~Tessarotto\corref{cor}}
\ead{fulvio.tessarotto@ts.infn.it}
%\emailAdd{fulvio.tessarotto@ts.infn.it}
\author[a]{Triloki}
\author[c]{A.~Valentini}
\author[d]{L.~Velardi}

\cortext[cor]{Corresponding author}

%\affiliation[a]{INFN Trieste  Trieste Italy}
%\affiliation[b]{University of Trieste and INFN Trieste, Trieste, Italy}
%\affiliation[c]{University Aldo Moro of Bari and INFN Bari, Bari, Italy}
%\affiliation[d]{CNR-ISTP and INFN  Bari, Bari, Italy}
%\affiliation[e]{Abdus Salam ICTP, Trieste, Italy and INFN Trieste, Trieste, Italy}
%\affiliation[f]{European Organization for Nuclear Research (CERN), CH-1211 Geneve 23, Switzerland}
%\affiliation[g]{Wrocław University of Science and Technology, Wybrzeże Wyspiańskiego 27, 50-370 Wrocław, Poland}
\address[a]{INFN Trieste  Trieste Italy}
\address[b]{University of Trieste and INFN Trieste, Trieste, Italy}
\address[c]{University Aldo Moro of Bari and INFN Bari, Bari, Italy}
\address[d]{CNR-ISTP and INFN  Bari, Bari, Italy}
\address[e]{Abdus Salam ICTP, Trieste, Italy and INFN Trieste, Trieste, Italy}
\address[f]{European Organization for Nuclear Research (CERN), CH-1211 Geneve 23, Switzerland}
\address[g]{Wrocław University of Science and Technology, Wybrzeże Wyspiańskiego 27, 50-370 Wrocław, Poland}

%\ead{Triloki@ts.infn.it}

\begin{abstract}
	
The next generation of gaseous photon detectors is requested to overcome the limitations of the available technology, in terms of resolution and robustness.
% The proposed new Electron-Ion Collider poses a technical and intellectual challenge for the detector design to accommodate the long-term diverse physics goals envisaged by the program. This requires a 4$\pi$ detector system capable of reconstructing the energy and momentum of final state particles with high precision. The Electron-Ion Collider also requires identification of particles of different masses over a wide momentum range.
%\par
%The Electron-Ion Collider requires identification of particles of different masses over a wide momentum range. A diverse spectrum of Particle IDentification detectors has been proposed. Of the four types of detectors for hadron identification, three are based on Ring Imaging Cherenkov Counter technologies, and one is realized by the Time of Flight method.
The quest for a novel photocathode, sensitive in the far vacuum ultra violet wavelength range and more robust than present ones, motivated an R\&D programme to explore nanodiamond based photoconverters, which represent the most promising alternative to cesium iodine.
A procedure for producing the novel photocathodes has been defined and applied on THGEMs samples.
%performed by a collaboration between INFN and CNR Bari and INFN Trieste.
 Systematic measurements of the photo emission in different Ar/CH$_{4}$ and  Ar/CO$_{2}$ gas mixtures with various types of nanodiamond powders have been performed. A comparative study of the response of THGEMs before and after coating demonstrated their full compatibility with the novel photocathodes.
%\par
%The progress of this R\&D programme and the results obtained so far by these exploratory studies are described.
  
\end{abstract}

\end{frontmatter}

\section{Introduction}
\label{sec:intro}
Gaseous Photon Detectors (PDs) have played an essential role in RICH applications and represent one of the best options for covering large areas with photosensitive detectors. They are not commercially available but they can be produced with a moderate investment; their operation is compatible with the presence of a magnetic field and they can offer a low material budget.
The first gaseous PDs operated with photo-ionizing vapours, but CsI photocathodes are currently used in almost all applications.
\par
CsI-based gaseous PDs have a high Photon Detection Efficiency (PDE) in the wavelength range below 200 nm. 
CsI is more robust than other photoconverting materials used in vacuum-based PDs but,
in spite of its widespread use and successful applications, it presents limitations due to its hygroscopic nature. A degradation in Quantum Efficiency (QE) appears in case of water vapour absorption~\cite{NIMA_695_2012_279} making the handling of CsI photocathodes a very delicate operation. QE degradation also appears after intense ion bombardment, when the integrated charge is $\sim$ 1~mC/cm$^{2}$~\cite{NIMA_574_2007_28} or larger. This limits the possibility to use CsI-based gaseous PDs in harsh environment.
%while in the visible range is effective photoconversion in gas is acivehas not been fully achieved yet.
The main source of ion bombardment is the ion avalanche produced in the multiplication process; the fraction of ions reaching the cathode depends on the detector architecture.
\par
In recent years, MPGD-based PDs with enhanced ion blocking capability have been %proposed~\cite{MPGD},
developed~\cite{ibf-blocking} and implemented for physics measurements.
The four MPGD-based PDs~\cite{PM18} of COMPASS RICH-1 represent
the most successful application of this technology: they have a hybrid architecture
consisting of two THGEM multiplication layers (the first one coated with CsI) and a
Micromegas. They cover a total active area of 1.4 m$^2$ and are in operation since 2016.
%\par 
%The MicroPattern Gaseous Detector (MPGD)-based Photon Detectors (PD) \cite{MPGD} have recently been demonstrated as effective devices~\cite{PM18} for the detection of single photon in  Cherenkov imaging counters. These PDs are composed of a hybrid structure, where two layers of THick GEM (THGEM) multipliers~\cite{thgem} are followed by a MICRO-MEsh GAseous Structure (MICROMEGAS)~\cite{mm} stage; the top layer of the first THGEM is coated with a reflective CsI PhotoCathode (PC).

The development of more robust photoconverters for gaseous PDs, for future RICH applications at new facilities, for the timing resolution frontier, etc. needs a dedicated effort. 

The EPIC collaboration ~\cite{EPIC} at the EIC~\cite{EIC}, a facility aimed to achieve understanding of Quantum ChromoDynamics (QCD), in particular in the elusive non-perturbative domain,
% and the answer to key questions, pending since long. Among them: the origin of nucleon mass and spin  and the properties of dense gluon systems.
requires efficient hadron identification in a wide particle momentum range, imposing the use of the gaseous RICH technique.
%including the challenging scope of hadron PID at high momenta, namely larger than $6$-$8~GeV/c$. A gaseous Ring Imaging CHerenkov (RICH) is the only possible choice for this specific task. The number of Cherenkov photons generated in a light radiator is limited. In spectrometer setups,  these number of photons is recovered by using long radiators.
The compact design of the EPIC Detector imposes limitations on the RICH radiator length. %requiring a dedicated strategy. In the far ultraviolet (UV) spectral region ($\sim120$~nm), the number of generated Cherenkov photons is larger, according to the Frank-Tamm distribution~\cite{Frank:1937fk}. This suggests the detection of
A possible solution could be the detection of photons in the very far UV range
by windowless gaseous PDs operating 
%. The standard fused-silica windows are opaque for wavelengths below 165~nm. Therefore, a windowless RICH is a potential option, implying the use of gaseous photon detectors operated 
with the radiator gas itself~\cite{windowless-RICH}, with high gain at high rate. This solution challenges the CsI performance limitations.
%\par 
%The MicroPattern Gaseous Detector (MPGD)-based Photon Detectors (PD) \cite{MPGD} have recently been demonstrated as effective devices~\cite{PM18} for the detection of single photon in  Cherenkov imaging counters. These PDs are composed of a hybrid structure, where two layers of THick GEM (THGEM) multipliers~\cite{thgem} are followed by a MICRO-MEsh GAseous Structure (MICROMEGAS)~\cite{mm} stage; the top layer of the first THGEM is coated with a reflective CsI PhotoCathode (PC).

\par
The quest for an alternative UV-sensitive photocathode overcoming these limitations is the main motivation for the R\&D programme discussed in this article. We first recall the novel procedure to produce Hydrogenated NanoDiamond (H-ND) based photocathodes and the recent results on the photon extraction efficiency in different Ar/CH$_{4}$ gas mixtures. We then illustrate the study of the compatibility between H-ND and THGEMs and present first results from the systematic comparison of the response of THGEMs as electron multipliers before and after the H-ND coating.
%THGEMs are robust gaseous electron multipliers based on GEM, principle scaling the geometrical parameters. 
%They are obtained via standard PCB drilling and etching processes. The 35~$\mu$m copper layer is coated with $\approx$5~$\mu$m of Ni, followed by 200~nm Au.
All THGEMs used for the studies reported in this article have an active area of $30\times30~$mm$^{2}$ and the same geometrical structure as COMPASS RICH-1 THGEMs: hole diameter of 0.4~mm, pitch of 0.8~mm, thickness of 470~$\mu$m and no rim around the holes.

%%%%%%%%%%%%%%%%%%%%%%%%%%%%%%%%%%%%%%%%%%%%%%

%\section{PhotoCathode (PC) based on Nanodiamond (ND) alternative to Caesium Iodide (CsI)}
%\section{Nanodiamond powder as a replacement of CsI PC}
%\section{Nanodiamond-based photocathodes as an alternative of CsI PC}
\section{Nanodiamond-based photocathodes preparation and first trials on THGEMs}
The high QE value of CsI photocathodes is related to its low electron affinity (0.1~eV) and wide band gap (6.2~eV) ~\cite{JAP_77_1995_2138}. The NanoDiamond (ND) particles have a comparable band gap of 5.5~eV and low electron affinity of 0.35-0.50~eV, and they exhibit chemical inertness and good radiation hardness: they represent a potential candidate for  photo-converter material more robust than CsI.
ND hydrogenation lowers the electron affinity to -1.27~eV,
%The negative electron affinity allows
favoring an efficient escape into vacuum of the generated photo electrons~\cite{NDRep-1}.
%without an energy barrier at the surface

%In literature H-ND is reported by Velardi et al., to have 22\% QE at 146 nm , while QE of CsI at 146 nm is reported by Rabus at al., to be
%~$\sim {45}\%$ at ${60}^{0}C$  and
%~$\sim {40}\%$ at ${25}^{0}C$ ~\cite{NDRep-1, NIMA_438_1999_94}.
\par
The standard procedure of hydrogenation of ND powder photocathodes is performed by using the MicroWave Plasma Enhanced Chemical Vapor Deposition (MWPECVD) technique at $\sim$ 800 $^{0}$C.
The hydrogenated powder is then cooled down to room temperature under high vacuum.
This procedure cannot be used for THGEMs which are made of fiberglass, and do not tolerate temperatures above 180~$^{0}$C.
%This limitation is overcome by the novel and low-cost technique developed at INFN Bari  ~\cite{coating-I, coating-II}. 
\par
A novel coating procedure, developed in Bari~\cite{coating-I, coating-II},
provides high QE~\cite{NDRep-1} and low-temperature ($\sim$ 120~$^{0}$C) deposition, allowing to cover different types of substrates, including standard THGEMs, with H-ND photoconverting layers.   
\par
The ND or H-ND powder is mixed with deionized water in one to one proportion and sonicated for 60 minutes. The solution is sprayed using a pressure atomizer with a LabVIEW 
software controlled system in pulses (shots) of 100 ms at about 3 Hz frequency.

\begin{figure}
	\begin{center}
	\includegraphics[width=0.65\linewidth]{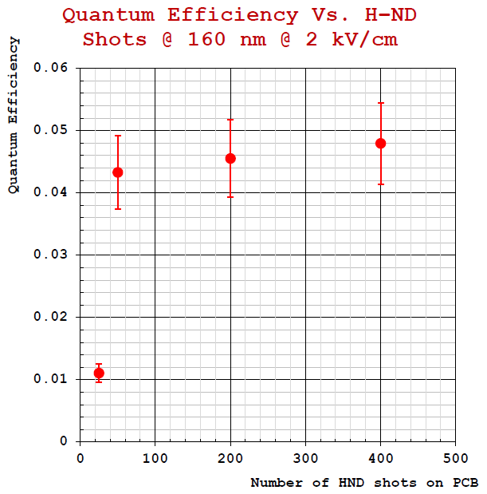}
	\vskip -0.5 em
	\caption{The QE of H-ND coated PCB substrates for 160 nm photons has been measured for samples coated with different quantity of H-ND powder, in units called shots. For amounts corresponding to 50 shots or more the QE is constant, indicating sufficient surface coverage.
	}
	\label{fig:ND_shots}
	\end{center}
\end{figure}

The substrates are kept at $\sim$150$~^0$C and rotated by magnetic stirrer; the distance between the atomizer nozzle and the substrate is $\sim$ 10 mm. We now deposit $\sim$ 250 shots/cm$^2$, according to the results of a previous study of photoemission as function of the number of shots, in which we observed that the photocurrent saturated above 50 shots/cm$^2$, as can be seen in Fig.~\ref{fig:ND_shots}.
%\vskip -2.0 em

After the coating a heat treatment (18 h in oven at 120 $^0$C) is applied: this treatment has been verified not to alter the QE response of the samples.

%A comparison of CsI and ND QE can  be extracted from  literature ~\cite{NDRep-1, NIMA_438_1999_94}, where, at 146 nm wavelength CsI photocathode is having more than 45\% QE while, for H-ND photocathode it is reported about 22\%.  

%\subsection{ND hydrogenation and coating procedure}
\par
Preliminary measurements of the photosensitivity of PCB substrates coated with ND and H-ND have been performed before fixing the protocol described above: while the standard PCB substrate samples were always offering promising results, some of the first coated THGEM samples could not be operated as electron multipliers due to severe losses in the electric strength after being coated with H-ND.
\par
A systematic study was then started to verify the compatibility of H-ND with THGEMs: a precise definition of the production and test procedures and an accurate characterization of the response of many THGEM samples before and after coating are ongoing.
In parallel, a test of the aging properties of H-ND coated samples~\cite{aging} was performed at CERN and showed that H-ND is at least ten times more robust than CsI against ion bombardment.
%The new samples are coated with 250 shots/cm$^2$. 
%%%%%%%%%%%%%%%%%%%%%%%%%%%%%%%%%

\section{Measurement of H-ND photocurrent in different gas mixtures}

\par
Three types of ND powders have been used in 2021 for the new study of both QE and THGEM compatibility. 
Two types have an average grain size of 250 nm and were produced by Diamonds \& Tools (D\&T) srl and Element 6 (E6); the third type, boron doped (BDD) produced by SOMEBETTER : ChangSha 3 Better Ultra-hard materials Co.,LTD has an average grain size of 500 nm. 

Test substrates have been coated with the three types of powders, both without hydrogenation and with hydrogenation, and their QE was measured for each of them, for different wavelengths, and different applied electric field values:
the results of this systematic study are reported in another article~\cite{NIMA_NDIP2022};
the hydrogenated D\&T ND powder provides the highest quantum efficiency for all wavelengths in vacuum.
%with fixed electric field (0.2 kV/cm)  applied over the surface in vacuum as shown in Fig.\ref{fig:qeallcomp}-Left. The ND and H-ND powders from D\&T provided the highest QE for all wavelengths and are shown separately in Fig.~\ref{fig:qeallcomp}-Right. 

%%%%%%%%%%%%%%%%%%%%%%
%\begin{figure}
%\begin{minipage}[c]{1\textwidth}
%    \includegraphics[width=\textwidth]{Quantum_efficiency_comparison.png}
%    \includegraphics[width=0.9\linewidth]{QE_comparison.png}
%\end{minipage}\hfill
%\begin{minipage}[c]{1\textwidth}
%        \caption{Left: QE as a function of wavelength for different types of ND and H-ND coated PCB substrates.}
%        \label{fig:qeallcomp}
%    \end{minipage}
%\end{figure}
%%%%%%%%%%%%%%%%%%%%%%

\par
\begin{figure}
	\begin{minipage}[c]{0.6\textwidth}
		\includegraphics[width=0.99\linewidth]{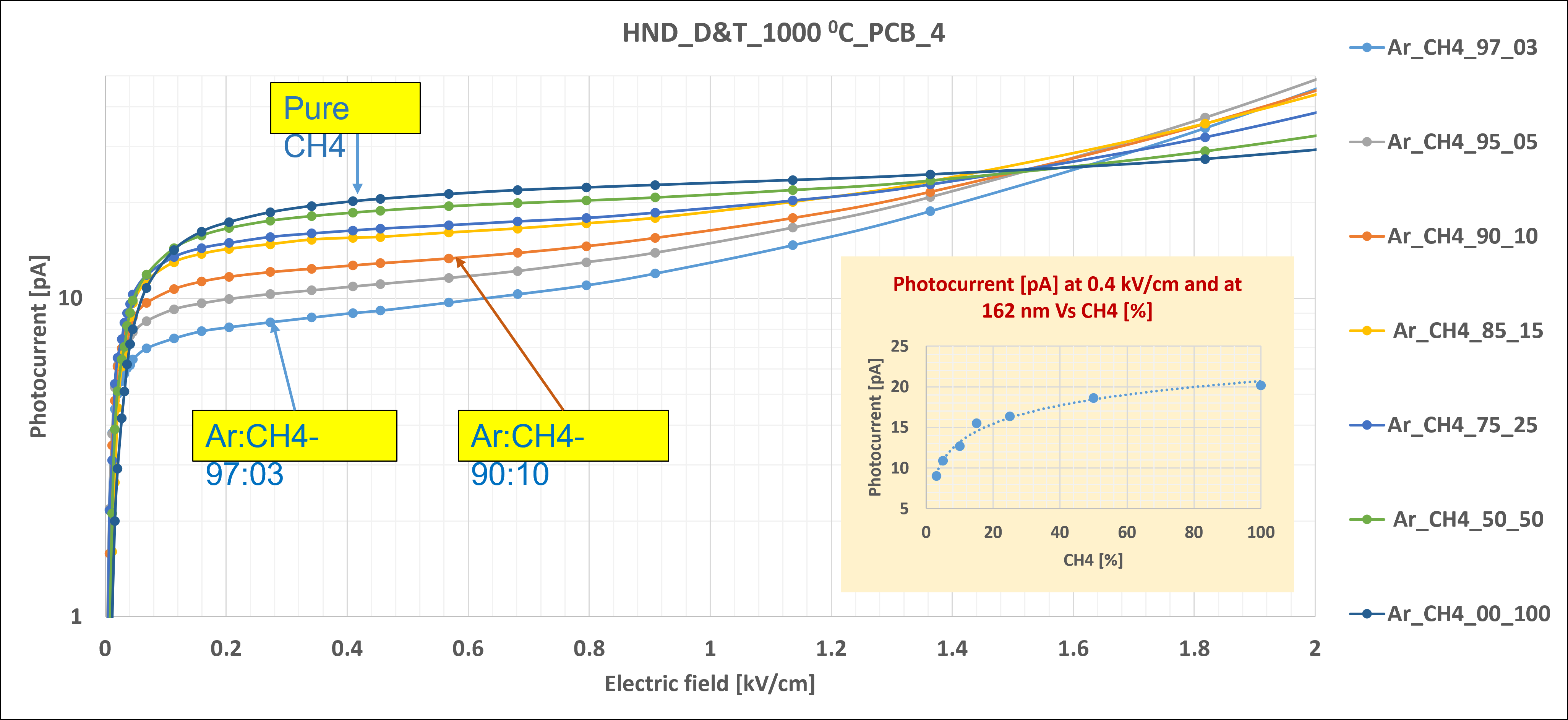}
	\end{minipage}\hfill
%	\begin{minipage}[c]{1\textwidth}
		\caption{Photocurrent as a function of applied electric field over the photo-cathode surface for different Ar/CH$_{4}$ gas mixtures at 160 nm wavelength. The photocurrent values for 0.4 kV/cm as function of the fraction of CH$_4$ are presented in the inserted colored canvas.}
		\label{fig:PhEmFull}
%	\end{minipage}
\end{figure}

Systematic measurements were also performed at the peak wavelength of 160 nm for different electric fields in different Ar/CH$_{4}$ gas mixtures. We present in Fig.~\ref{fig:PhEmFull}
the measurement of photocurrent as funtion of applied electric field for seven gas mixtures.
For electric field value lower than 1 kV/cm, the measured photo current in argon rich gas mixtures is lower compared to that of CH$_{4}$ rich mixtures, due to the high photon back scattering cross section of argon.
\par
For mixtures with low percentage of CH$_{4}$, the effect of gas multiplication is visible for electric field values larger than 0.5 kV/cm; above $\sim$ 1.5 kV/cm the mixtures rich in argon provide larger total measured photocurrent values compared to CH$_{4}$ rich ones, because the gas multiplication effect becomes dominant.

\par
The photocurrent values for 0.4 kV/cm as function of the fraction of CH$_4$ are presented in the colored canvas inserted in Fig\ref{fig:PhEmFull}: With respect to the
%photocurrent provided by the
pure CH$_{4}$ gas case, the Ar/CH$_{4}$ 50/50 mixture shows a 9\% lower photocurrent value, the Ar/CH$_{4}$ 75/25 a 24\% lower value and the Ar/CH$_{4}$ 97/03 a 55\% lower value.

%%%%%%%%%%%%%%%%%%%%%%

%The value of the electric field used is given in revised manuscript.

\section{Study of H-ND compatiblity with THGEMs}

%%%%%%%%%%%%%%%%%%%%%%%%%%%%%%%%%%%%%%%%%%%%%%
%\subsection{Characterization before Coating}

%\par
%THGEMs are robust gaseous electron multipliers based on GEM, principle scaling the geometrical parameters. 
%They are obtained via standard PCB drilling and etching processes. The 35~$\mu$m copper layer is coated with $\approx$5~$\mu$m of Ni, followed by 200~nm Au. The THGEMs used for our studies have an active area of $30\times30~mm^{2}$ with a  hole diameter of  0.4~mm, a pitch of 0.8~mm, a thickness of 470~$\mu$m and no rim around the holes.
\par

The characterization of each THGEM is performed in a simple test setup including
%sketched in figure  ~\ref{fig:Schematic_of_Detector_setup}. 
a plane of drift wires above the THGEM and a segmented readout anode plane below it, both properly biased, providing the drift and induction field respectively. The detector is operated with various gas mixtures, all including argon. The electrons from $^{55}$Fe X-Rays, converted by argon, are collected and multiplied in the THGEM holes. The electron avalanche generated in the multiplication process, while drifting towards the anode, induces 
the detected signal. 

%%%%%%%%%%%%%%%%%%%%%%
\begin{figure}
	\begin{flushright}
%\begin{minipage}[c]{1\textwidth}
    \includegraphics[width=0.8\linewidth]{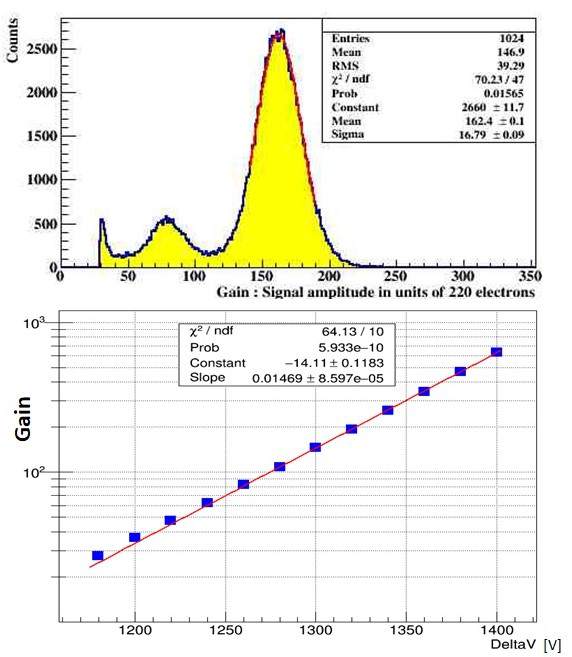}
%    \includegraphics[width=\textwidth]{Schematic_of_Detector_setup.jpg}
%\end{minipage}\hfill
%\begin{minipage}[c]{0.98\textwidth}
        \caption{Top: typical $^{55}$Fe X-ray spectrum obtained in Ar/CO$_{2}$~70\%/30\% gas mixture when the applied voltages at drift, top and bottom of THGEM are -2520 V, -1720 V and -500 V, respectively, while the anode is at ground voltage. Bottom: gain dependence of the THGEM as function of the applied voltage.}
        \label{fig:Schematic_of_Detector_setup}
%    \end{minipage}
	\end{flushright}
\end{figure}
%%%%%%%%%%%%%%%%%%%%%%%%

All THGEMs have been characterized using a Ar/CO$_2$ 70/30 gas mixture and a  $^{55}$Fe  X-ray source at INFN Trieste before applying the spray procedures, with the goal of performing comparative studies after coating them with VUV sensitive films. Environmental conditions (gas pressure and temperature) were registered and used for the corrected effective gain determination.
The THGEM gain response has a variation in time related to the effect of charging up of
the dielectric surfaces: a gain evolution study of about three days has been performed for each THGEM sample.
The same characterization has been repeated after coating.
\par 
 A typical $^{55}$Fe X-ray spectrum obtained in  Ar/CO$_{2}$ 70\%/30\% gas  mixture is shown in the top panel of Fig.~\ref{fig:Schematic_of_Detector_setup}. The bottom panel shows the gain dependence of the THGEM as function of the applied bias voltage.

%%%%%%%%%%%%%%%%%%%%%%
%\begin{figure}
%\begin{minipage}[c]{1\textwidth}
%    \includegraphics[width=\textwidth]{Detector_Images.png}
%\end{minipage}\hfill
%\begin{minipage}[c]{0.98\textwidth}
%        \caption{ (A) Au\_PCB of 1 inch diameter substrate used for the QE measurement. (B)  Uncoated THGEM of active area 30~mm$\times$30~mm. (C) Half uncoated and half coated THGEM, mounted into the test chamber and zoomed view of the both coated (D) and uncoated (E) part. (F) test chamber with readout pad where the THGEMs are tested. (G) The test chamber after installation of a THGEM, illuminated by an  ${}^{55}Fe$ X-ray source.}
%        \label{fig:Detector_Images}
%    \end{minipage}
%\end{figure}
%%%%%%%%%%%%%%%%%%%%%%%%

Each THGEM sample has been coated in Bari either with as-received ND powder, or with H-ND. 
To allow easier comparison, for some THGEMs only half of the active surface has been coated,
leaving the other half uncoated. The same coating procedure (and the same amount of ND per unit surface) has been applied to the PCB test substrates and to the THGEMs.  
 %\par
 %Images of the coated substrates and the setup for the characterization are provided in Fig.~{\ref{fig:Detector_Images}}.

%%%%%%%%%%%%%%%%%%%%%%%%%%

\begin{figure}
	\begin{center}
\includegraphics[width=0.70\linewidth]{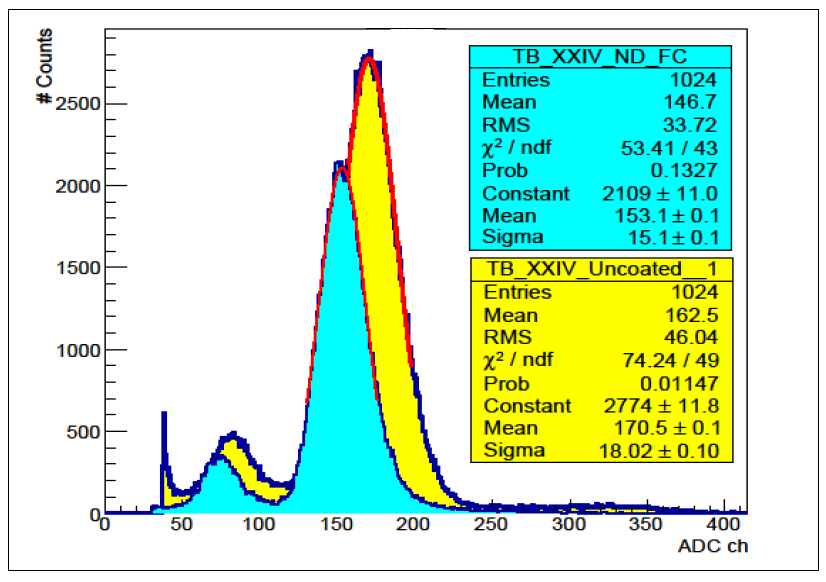}
\includegraphics[width=0.68\linewidth]{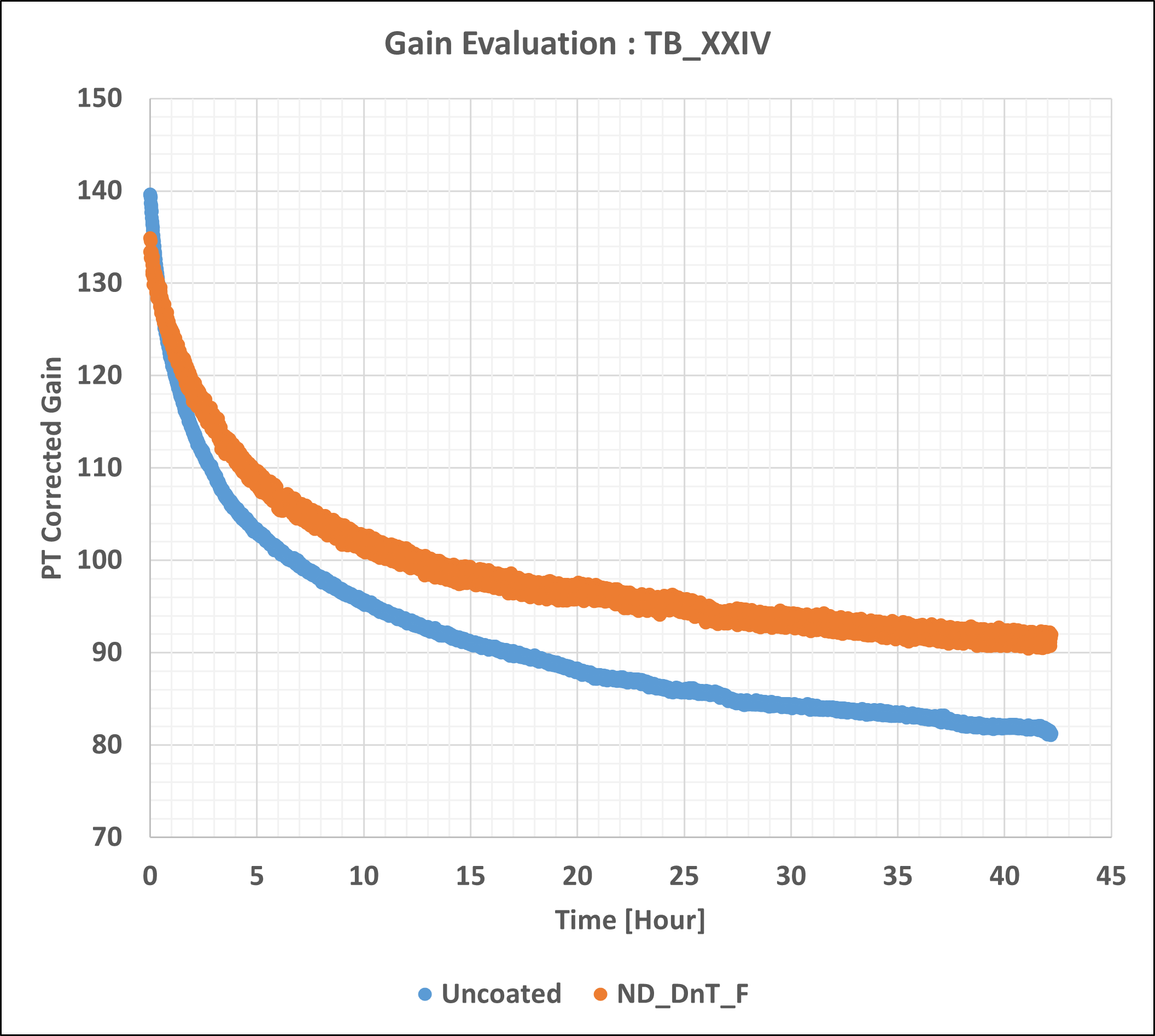}
\caption{Top: typical ${}^{55}Fe$ X-ray spectra obtained from a same THGEM-XXIV  before coating and after fully coated ND powder in  Ar/CO$_{2}$ 70\%/30\% gas  mixture.  The voltages applied  to drift, top and bottom of the THGEM electrodes are -2250 V, -1750 V and -500 V respectively; the anode is kept at ground voltage.
Bottom: gain evolution due to charging-up of the THGEM measured before (blue) and after (orange) coating; corrections for pressure and temperature have been applied.
}
\label{fig:ND_THGEM_Spectra}
	\end{center}
\end{figure}

\par 
Some of the fully characterized THGEMs and their respective coatings are listed below: 

\begin{table}[!ht]
    \centering
    \begin{tabular}{|l|l|l|l|}
    \hline
        {\small Uncoated THGEMs}
        	  & {\small Max $\Delta$V} & {\small Coated THGEMs}& {\small Max $\Delta$V}   \\ \hline
        TB\_XV & 1325 & H-ND\_BDD\_H & 1325 \\ \hline
        TB\_XVI & 1375 & ND\_E6\_H & 1400 \\ \hline
        TB\_XVII & 1350 & ND\_BDD\_H & 1325 \\ \hline
        TB\_XXIV & 1375 & ND\_D\&T\_F & 1325 \\ \hline
    \end{tabular}
    \caption{Maximum stable voltage of few uncoated and coated THGEMs in  Ar/CO$_{2}$ 70\%/30\% gas mixture.}
    \label{tab:TableDelVThGEM}
\end{table}

The first column of Tab.\ref{tab:TableDelVThGEM} shows the THGEM labels, the second one the maximum voltage bias at which the THGEM was operating stably; in the third column the coating material for each THGEM are listed and in the fourth the maximum stable voltage after coating is provided. The electrical stability of THGEMs does not seem to be systematically affected by the coating layers. The variations are compatible with the changes in environmental conditions.
\par
Typical examples of amplitude distributions for $^{55}$Fe signals from an uncoated (yellow) and ND coated (light blue) THGEM are shown superimposed in the top part of Fig.~\ref{fig:ND_THGEM_Spectra}. The amplitude spectra are presented before application of the correction for the different pressure and temperature conditions. The effective gain is extracted from the mean of the Gaussian fit of the main peak of the spectrum.
To study the charging-up response, a continuous set of short ($\sim$ 30 sec.) acquisition runs
over few days is performed. In the bottom part of Fig.~\ref{fig:ND_THGEM_Spectra} the charging-up curve is presented for the same THGEM before (blue) and after (orange) coating; for this comparison the effective gain values have been corrected for pressure and temperature.
The difference seen in the charging up variation rate is partly due to the change in the $^{55}$Fe source activity. 
\par
\begin{figure}
		\begin{center}
	\includegraphics[width=.92\linewidth]{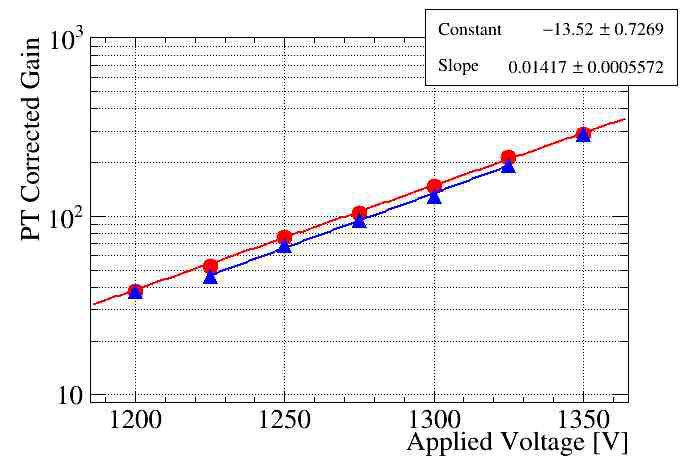}
	\caption {Comparison between corrected effective gain versus applied voltage across the THGEM electrodes before (blue) points and after (red points) coating.}
	\label{fig:HV_scan}
		\end{center}
\end{figure}

A typical plot of the effective gain values, corrected for pressure and temperature, as a function of the bias voltage applied to the THGEMs is presented in Fig.~\ref{fig:HV_scan}.
The blue points represent the corrected gain values before coating; the red ones the gain values after coationg.
%All tested THGEM samples present the same conclusion:
The characterization of a THGEM without and with various ND coatings provides almost identical results. The gain values are found to be similar in all cases: the THGEM response as electron multiplier is not significantly modified by any of the ND and H-ND coatings.
\par
To achieve a more accurate evaluation of possible small effects, more THGEM samples will be
tested. A newly built, complete photon detector prototype is now available and will be used in a campaign of tests for the next step in the validation of the "MPGD + H-ND technology" for the detection of single VUV photons for RICH applications.
Other fields can potentially benefit from the development of this innovative technique.  

%%%%%%%%%%%%%%%%%%%%%%%%%%%%%%%%%%%%%%%
%%%%%%%%%%%%%%%%%%%%%%%%%%%%%%%%%%%%%%%

%%%%%%%%%%%%%%%%%%%%%%

\section{Conclusion}
The ongoing R\&D programme to explore the H-ND photocathode response in the VUV range for future gaseous PD application has provided encouraging results.
Photoemission current measurements in the VUV range in vacuum and various gas mixtures were performed for ND and H-ND samples. 
Photoemission currents measured at 160 nm as a function of the electric field, in various Ar/CH$_{4}$ compositions, allowed to distinguish between two different voltage regimes. At lower field values the important role of CH$_{4}$ in reducing the photoelectron backscattering is highlighted, providing a factor of two in photoconversion probability between Ar-rich mixtures and pure CH$_{4}$.
 At higher voltages the gas multiplication dominates, in particular for the gas mixtures with large argon fraction. 
\par 
THGEM samples coated with different types of photosensitive layers of ND and H-ND, from D\&T, E6 and BDD  have been studied with extensive characterization. 
The stuedied THGEMs sustain more than 1325 V bias across top and bottom surfaces.
The response of THGEMs as electron multipliers is not significantly modified by the ND or H-ND coating.
The results of this first exploratory study, suggesting a full compatibility of H-ND with THGEMs encourage to proceed in the development of gaseous PDs with a photocathode overcoming the limitations of CsI.

%%%%%%%%%%%%%%%%%%%%%%%%%%%%%%%%%%%%%%%%%
\section{Acknowledgment}
This R\&D activity is partially supported by
\begin{itemize}
\item 
EU Horizon 2020 research and innovation programme, STRONG-2020 project, under grant agreement No ;
\item 
the Program Detector Generic R\&D for an Electron Ion Collider by Brookhaven National Laboratory, in association with Jefferson Lab and the DOE Office of Nuclear Physics.
\end{itemize}
%Triloki acknowledge conference travel support from the international center for theoretical physics (ICTP), Italy.
%%%%%%%%%%%%%%%%%%%%%%%%%%%%%%%%%%%%%%%%%

\end{document}